\documentclass[letterpaper, 10 pt, conference]{ieeeconf}  

\IEEEoverridecommandlockouts                              

\usepackage{color}
\usepackage{graphicx}
\usepackage{subfigure}
\usepackage{cite}
\usepackage{amsmath}
\usepackage{amsfonts}
\usepackage{enumerate}

\graphicspath{{images/}}

\newcommand{\MATE}{{MATE}}

\title{\LARGE \bf \MATE{} Robots Simplifying My Work:\\The Benefits and Socioethical Implications}

\author{Valeria Villani$^{1}$, Lorenzo Sabattini$^{1}$, Julia N. Czerniak$^{2}$, Alexander Mertens$^{2}$ and Cesare Fantuzzi$^{1}$
	\thanks{$^{1}$V. Villani, L. Sabattini and C. Fantuzzi are with the Department of Sciences and Methods for Engineering (DISMI), University of Modena and Reggio Emilia, Reggio Emilia, Italy
		{\tt\small \{valeria.villani, lorenzo.sabattini, cesare.fantuzzi\}@unimore.it}}%
	\thanks{$^{2}$J. N. Czerniak and A. Mertens are with the Institute of Industrial Engineering and Ergonomics, RWTH Aachen University, Aachen, Germany
		{\tt\small \{j.czerniak, a.mertens\}@iaw.rwth-aachen.de}}
}

\begin{document}

\maketitle
\thispagestyle{empty}
\pagestyle{empty}



\begin{figure}
	\centering
	\includegraphics[width=\columnwidth]{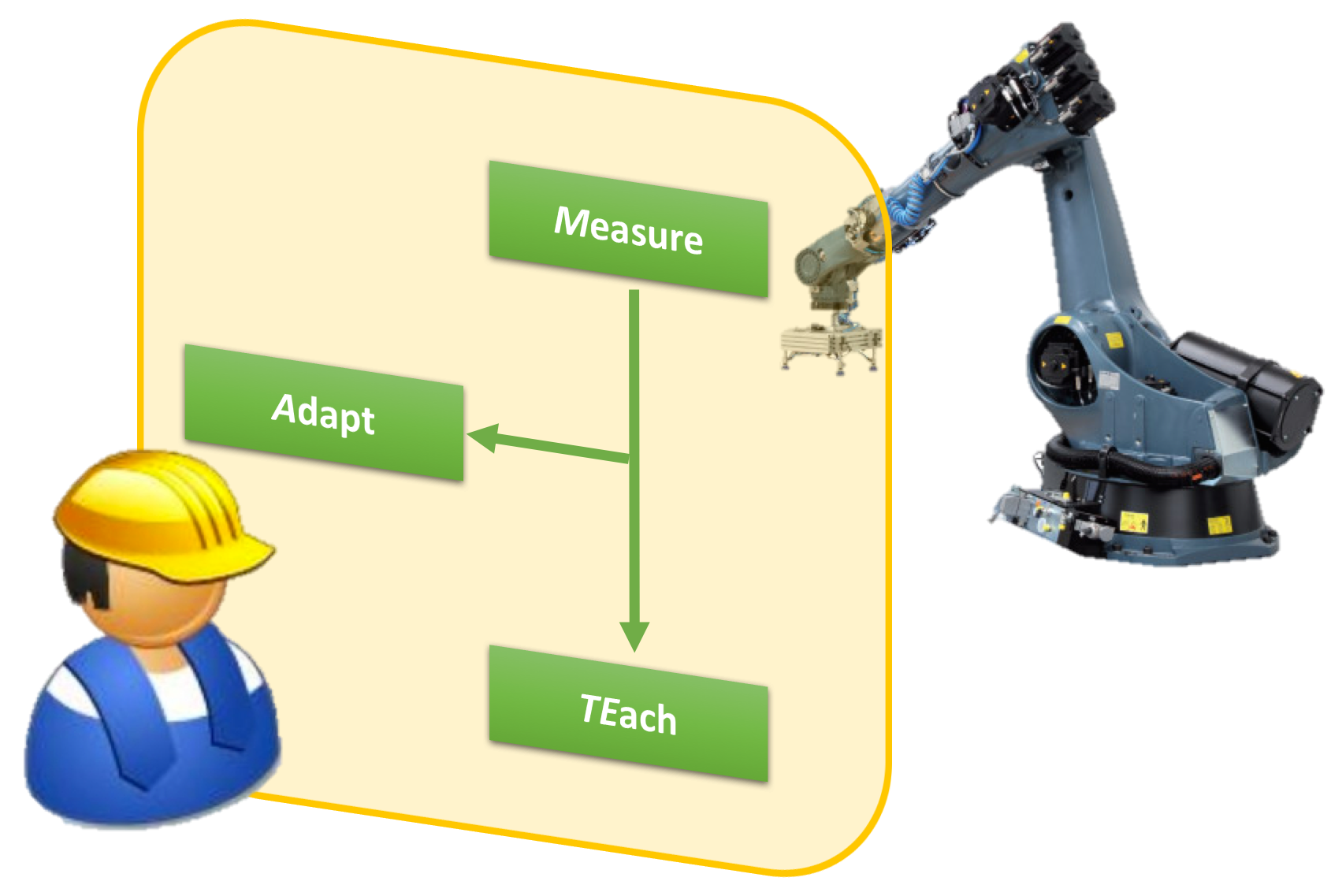}
	\caption{\label{fig:overview_inclusive}A schematic view of a \MATE{} system.}
\end{figure}

With the increasing complexity of modern industrial automatic and robotic systems, a burden is placed on system operators, who are required to supervise and interact with very complex systems, typically under difficult and stressful conditions. To overcome these challenges, it is necessary to adopt a responsible approach based on an anthropocentric design methodology so that machines adapt to human capabilities rather than vice versa. In this article, we consider an integrated methodological design approach, referred to as \emph{measure, adapt, and teach} (\emph{MATE}), which consists of devising complex automatic or robotic solutions that measure the current operator's status and adapting the interaction accordingly, while providing him or her with the necessary skills and expertise to improve the interaction. A MATE system, shown in Figure~\ref{fig:overview_inclusive}, endeavors to be usable for all users, thus meeting the principles of inclusive design. However, the use of such a MATE system calls to attention several ethical and social implications, which are discussed in this article. Additionally, a discussion about which factors in the organization of companies are critical with respect to the introduction of a MATE system is presented.

The increasing complexity of industrial automatic and robotic production systems is a result of industry competitiveness and the need to comply with market demands. As a consequence, along with this progress comes the need for laborers to acquire more advanced skills to operate such systems. Furthermore, they must also endure challenging work conditions, such as noisy environments, tight schedules, the fear of job loss, and/or psychological pressure due to the presence of supervisors. Such strenuous conditions are amplified when vulnerable users, such as those cognitively or physically impaired as well as elderly and low-educated operators, are involved in the interaction. In typical operative scenarios, these classes of workers are barred from job positions that necessitate the meticulous attention to detail required to interact with a robot or within a complex factory plant. Alternatively, in the case that these workers are granted any such occupations, their responsibilities and duties are severely limited. To invert such a policy, complex product systems need to be simplified. The adoption of a MATE system will enable such a goal, as the system strives to be understandable and meet the user's needs, to accomplish the desired tasks, and to provide an interaction that is a positive, enjoyable experience for the user \cite{Norman_2013}.

In the context of industrial production, this means reversing the current belief that humans must learn how the machine works to the future scenario where the machine adapts to the human capability by accommodating time and features \cite{Skripcak_2013,Nachreiner_2006}.

Interaction systems used in industrial environments do not provide the possibility of controlling the displayed information or its form; while the human operator is flexible and adaptable, the system is not. In particular, the control systems applied to industrial processes typically respond in a specified way without regard to whether the flow of information is low or high or whether the level of expertise of the user is good or bad \cite{Viano_2000_short}. The human operator, then, is typically the only one who adapts his or her behavior based on the situation. The operator must be sufficiently flexible to cope with common activities and unpredictable and/or dangerous situations. This can cause significant difficulties for the operators, resulting in a decreased level of job performance and satisfaction, as well as the prevention of less-skilled operators from using complex systems. To overcome these limitations and change the design approach, it becomes advantageous to include the user in the feedback loop of the system and his or her interaction with it; this instantaneously modulates the interaction based on the operator's current psychophysical status. This addresses the physical and cognitive overload produced by the working task and uses the information to adapt the interaction accordingly, relieving the user's stress and anxiety.

\subsection{MATE Approach}

A first attempt to implement this new paradigm is achieved by affective computing and robotics, which rely on measuring the user's physiological parameters that are related to mental strain and adapting the interaction with the robotic system accordingly \cite{Picard_1997, Landi_2017_MECH}. Extending the concept of affective computing, we aim to introduce an integrated methodological approach for the anthropocentric design of human-machine and robot systems. This approach, called MATE, relies on three important attributes: the measurement of human capabilities and skills, the adaptation of the interaction system, and the training and support for less-skilled users. The rationale behind the MATE approach is not only that the current operator's status is measured to adapt the interaction accordingly, but also that he or she learns the necessary skills and acquires the expertise essential to improve the interaction. Moreover, a more tailored adaptation of the interaction tasks and training support can be achieved by thoroughly characterizing the worker in terms of work skills, perceptive capabilities, and cognitive capabilities in addition to mood and affect, thereby extending the measurements related to affective computing. While the immediate goal of a system based on the MATE methodology is to improve the performance of operators while interacting with the system, the ultimate goal is to enhance the skills of vulnerable operators (i.e., the elderly, the cognitively or physically impaired, and low-educated operators) to attenuate their adverse conditions. Given its features, a MATE system is intended to be easy to use and usable for all operators, thus meeting the requirements of inclusive design \cite{universal_design,Stephanidis_2001,Abascal_2005}.

A first example of a MATE approach is being developed in the framework of the European project INCLUSIVE \cite{Villani_2017_ETFA, INCLUSIVE_url}. Its goal is to devise a methodology for the design of complex interaction systems that profile the user by measuring his or her cognitive and perceptive capabilities (offline) and sustainable cognitive burden (online), and then adapt the presentation of information and the interaction accordingly, providing adequate teaching support.

\subsection{Social and Ethical Impacts}

While most of the research on the use of collaborative solutions has focused on factors that have a direct influence on collaboration (e.g., motion and safety), limited research has focused on the anthropological impact. The social impact of the introduction of industrial and collaborative robots on factory workers is discussed in \cite{Sauppe_2015} and \cite{Elprama_2017}. With the introduction of robots in factories, social problems (i.e., unemployment resulting from the loss of jobs) are observed in \cite{Veruggio_2016} and discussed in interviews with factory workers in \cite{Elprama_2017}. Thus, the goal of having robots that support, rather than supplant, people in workplaces, as shown in \cite{Nevejans_2016} and \cite{Report_EU_2015}, has not been reached yet. Moreover, the use of robots and the topic of human–robot interaction give rise to several anthropological and ethical issues, which become more pronounced when vulnerable users are involved. To address this concern, frameworks for incorporating ethics into the design of robots \cite{vanWynsberghe_2016} that originated from care robots have been extended to service robots. However, considering the rapid growth of industrial collaborative robots, such frameworks should be properly adapted to account for industrial working scenarios. The term roboethics, which refers to a set of tools used to promote the development of robotics—while limiting its misuse and harm done to humans—has recently been introduced \cite{Veruggio_2016}. Specifically, roboethics addresses the ethical issues related to the use of robots, which are meant as intelligent devices intended and designed for making decisions on behalf of humans without conflicting with the will of humans nor replacing humans. This should not be confused with the coding of ethics or morals into robotic systems to make robots conscious beings \cite{Nevejans_2016}, which is beyond the scope of MATE systems.

When considering a MATE-robotic or MATE-automated system, e.g., the INCLUSIVE one, that adapts to its users to facilitate the interaction, social and ethical implications cannot be neglected (particularly, but not exclusively, when users with special needs are involved). Since the proper adaptation of the system to this classification of user relies on the accurate assessment of the user, the risk of stigmatization must be carefully handled.

\section{Anthropocentric Analysis of Requirements}\label{sec:requirements}

\begin{figure}
	\centering
	\includegraphics[width=1\columnwidth]{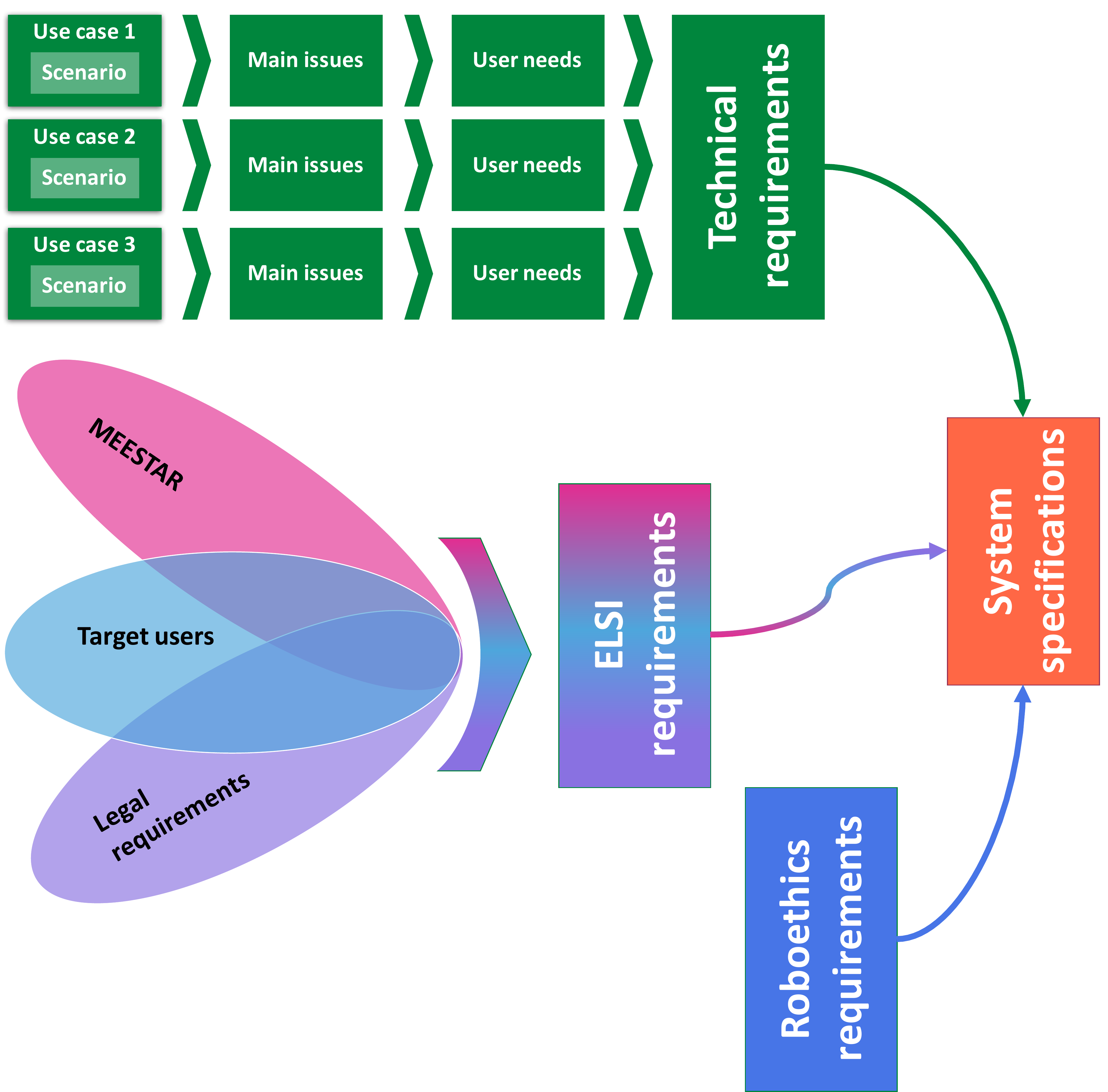}
	\caption{\label{fig:system_requirements}The process for defining technical, ELSI, and roboethics requirements for the design of human-centered complex automatic systems. MEESTAR: a model for the ethical evaluation of sociotechnical arrangements.}
\end{figure}

As with any complex system, the development of a MATE system requires defining specifications that provide a description of the main characteristics that the system must fulfill. Following a standard system engineering approach \cite{leonard1999}, these specifications are derived from a set of requirements that describe the desired behavior of the system from the user's perspective.

While system specifications depend on the specific application under consideration, the requirements can be defined from a universal point of view (for MATE systems, in general). Derived from an anthropocentric approach, these requirements focus on the user's needs.

Technical requirements were first considered, but they were complemented by additional necessary requirements regarding social and ethical implications of the system. Figure \ref{fig:system_requirements} demonstrates how the different classes of requirements are derived. All of the requirements, which have general validity, contribute to defining the technical specifications of a given system. Table~\ref{tab:design_recommendations} summarizes the proposed design recommendations in terms of technical requirements; requirements for the ethical, social, and legal implications (ELSI); and roboethics.

\begin{table*}
	\centering
	\caption{\label{tab:design_recommendations}Design recommendations for a \MATE{} system: technical and ELSI  requirements for the inclusion of vulnerable users and principles proposed by debate on roboethics~\cite{Nevejans_2016}.}
	\includegraphics[width=1.8\columnwidth]{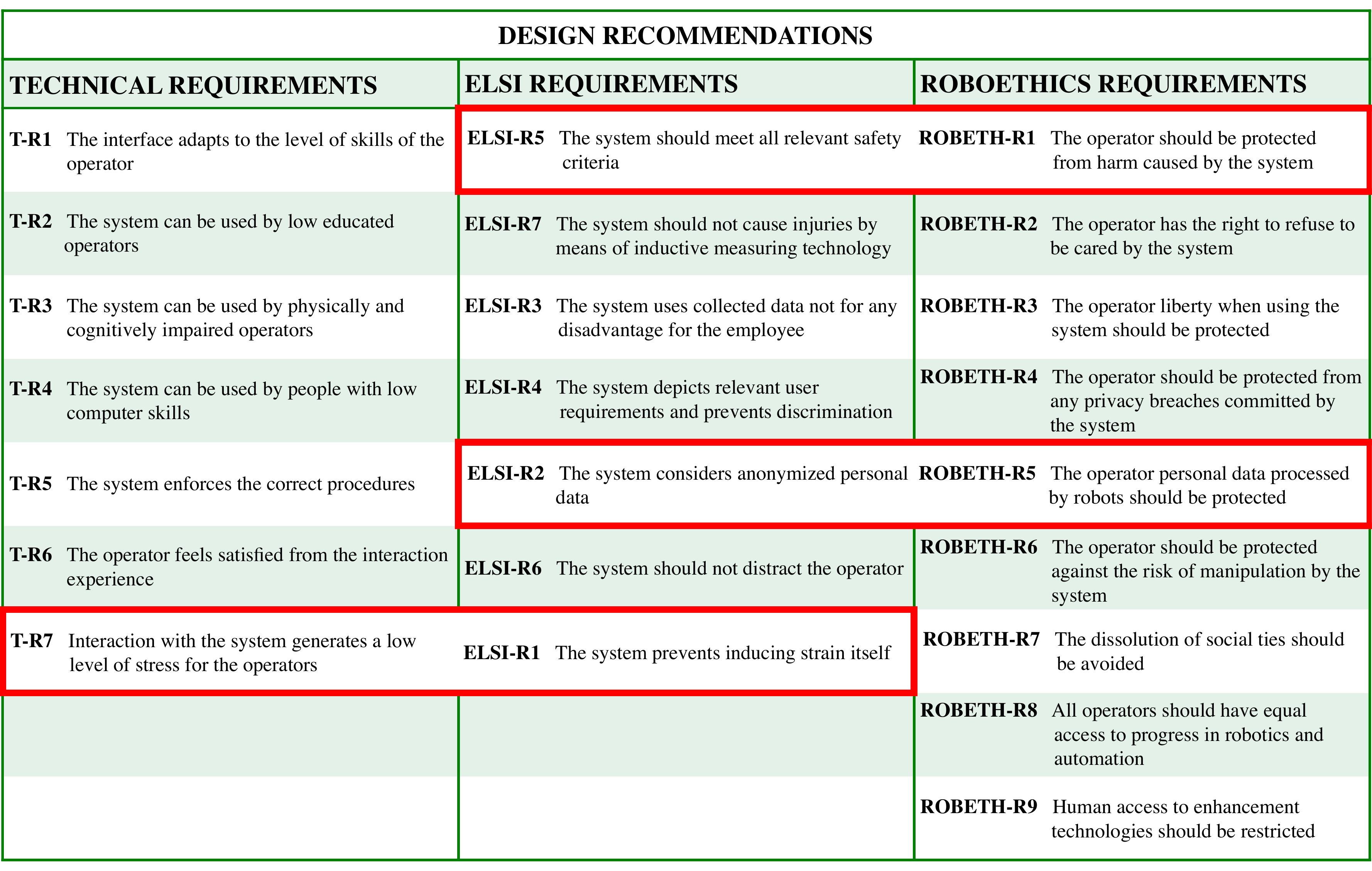}
\end{table*}

\subsection{Technical Requirements}
To obtain the design requirements, we formed a thorough understanding of the users and analyzed their needs. We focused on three specific industrial case studies that represent a broad area of interest in Europe:

\begin{itemize}[\textit{Use case} \itshape1\upshape:]
	\item \emph{use case 1}: machinery for woodworking, typically used in small companies run by elderly artisans
	\item \emph{use case 2}: robotic solutions to automatize the assembly of appliances, currently done manually
	\item \emph{use case 3}: automatic bottling machines used in industrial plants.
\end{itemize}

Such use cases have been analyzed in detail, with specific attention given to precise working scenarios and those tasks involving interaction between the human operator and the machine \cite{Sabattini_2017_CASE}. To understand the limitations and pitfalls that may reduce an operator's work performance, his or her current interaction with standard automation and robotics systems was observed \cite{Sabattini_2017_CASE}. Particular attention was paid to vulnerable users, since they are the most directly affected by the complexity of this interaction. For the purpose of these use cases, we consider vulnerable users not only those who have physical or cognitive impairments, but also those who are elderly and those with less education who may not be familiar with advanced technology and computerized solutions. These users are considered vulnerable because they are more likely to lose their jobs and they are less likely to be retrained or reemployed. This results from their limited ability to effectively utilize complex, modern computer-aided manufacturing equipment, as well as from impairments that restrict the performance of physically demanding tasks.

A list of users' needs, with respect to current human-machine systems, is summarized in Table \ref{tab:users_needs}. These needs highlight the fact that current interaction modality involves significant effort from users, particularly from vulnerable ones, which often results in additional stress for workers. A simplification of this interaction, by means of guidance in procedural tasks and easy-to-use interaction modes, must be achieved. The first set of requirements derived from these users' needs is listed in the left column of Table~\ref{tab:design_recommendations} and translates users' needs to technical hints for design to improve traditional approaches to interaction.

Note that these technical requirements are applicable to any inclusive interaction system and do not depend on the previously described use cases, though they have been derived from specific users' needs that are related to the use cases described in this article. Additionally, these users' needs have been developed as additional requirements to overcome the limitations of traditional systems, and they are not the only technical requirements to consider in the system's design (e.g., task-specific requirements and safety requirements).

\subsection{ELSI Requirements}\label{subsec_ELS_requirements}
In addition to the technical requirements obtained from the use cases, other design recommendations are derived from the analysis of ELSIs related to the use of an adaptive MATE human–machine system. This analysis helps to define the ELSI requirements. To develop an ELSI concept related to the design of smart interaction systems for automated production machines, we have combined a model for the ethical evaluation of sociotechnical arrangements (MEESTAR) \cite{Manzeschke_2015}, which considers ethical problems, and the context-specific legal issues for occupational systems, as well as societal needs of the vulnerable users \cite{Sabattini_2017_CASE}.

\begin{figure}
	\centering
	\includegraphics[width=.7\columnwidth]{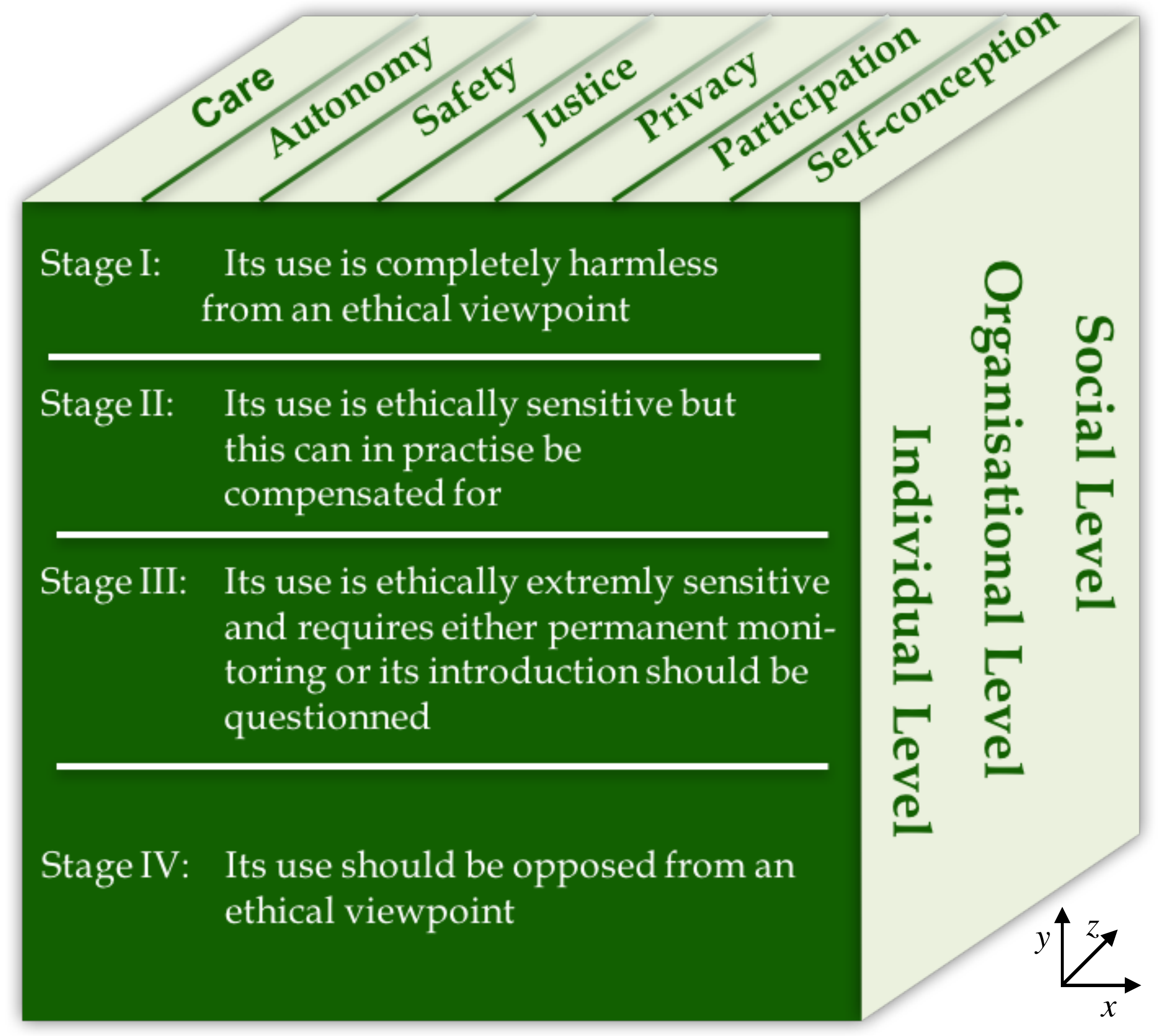}
	\caption{\label{fig:MEESTAR}A 3-D MEESTAR approach for the assessment of ethical issues related to the use of technology. x axis: dimensions of ethical evaluation; y axis: stages of ethical evaluation; z axis: levels of ethical evaluation \cite{Manzeschke_2015}.}
	\vspace{-.2cm}
\end{figure}

The MEESTAR approach is a three-dimensional (3-D) evaluation tool that guides users to ethically reflect upon and form judgments relative to the use of the appropriate systems. MEESTAR requires the system to cause little or no harm to the user and attempts to identify ethically problematic effects in a structured way to develop appropriate solutions. As shown in Figure \ref{fig:MEESTAR}, it divides this analysis into three main components and, when considered jointly, allows for a comprehensive ethical evaluation of a sociotechnical arrangement. The first dimension looks at ethical core values, i.e., care, autonomy, safety, justice, privacy, participation, and self-conception. The second dimension considers three levels of observation, i.e., individual, organizational, and societal, and it includes the perspectives of those who use the system and the people in close proximity to the user. The third dimension identifies four stages of ethical sensitivity, ranging from ``completely harmless'' to ``should be opposed from an ethical viewpoint.'' A detailed description of the MEESTAR approach is found in \cite{Manzeschke_2015}.

With regard to the legal implications, we consider data protection, safety, and health at work, as addressed by the European Union (EU) regulations, i.e., the Machinery Directive 2006/42/EC, the Council Directive 89/391/EEC, and the General Data Protection Regulation 2016/679. For the previously discussed vulnerable users, whose presumed limited perception skills and cognitive and motor skills would otherwise hinder their effectiveness, specific social requirements should be implemented to help ensure their equal treatment and integration into the work process.

\subsection{Roboethics Requirements}
The ELSI requirements previously presented do not provide a complete analysis of the design recommendations from an ethical point of view. Since we are dealing with vulnerable users and considering a system that measures human capabilities and performance, ethical issues are of specific concern and must be dealt with in detail. Additional ethical principles for the design, production, and use of robots are outlined in the framework of roboethics \cite{Nevejans_2016}. Although they are quite general, such roboethics requirements are well-suited to the MATE methodology and suggest further issues that are not covered by ELSI analysis. The requirements proposed by roboethics are reported in the last column of Table~\ref{tab:design_recommendations}. Two roboethics requirements (ROBETH-R1 and ROBETH-R5) overlap with ELSI principles (ELSI-R5 and ELSI-R2, respectively) and focus on the importance of the system not causing harm to the operator, as well as the use of anonymous personal data.

Roboethics acknowledges that the worker should always maintain control of the interaction (ROBETH-R2 and ROBETH-R3), dismissing it whenever he or she does not feel comfortable with the system. Thus, the user of a MATE system should have complete autonomy; for example, if a user experiences reluctance interacting with a MATE system, he or she should be free to use standard interaction approaches, though this choice would result in a decreased level of performance.

Roboethics requirement ROBETH-R6 focuses on one of the biggest drawbacks of using MATE systems in industrial applications: the risk of stigmatization that could arise from the use of a system that measures a worker's skills and performance. This may be a sensitive issue for the worker, who may feel embarrassment from the assessment, and for his or her supervisor or employer. Although this risk is partly mitigated by precautions put in place for protecting data, care must be taken when considering MATE approaches.

While social relationships in the workplace can be undermined by the introduction of robots, the MATE system only minimally aggravates this situation. Furthermore, the only decline that is observed in social ties among colleagues is due to less support being necessary for the user of a MATE system, since the system acts as support for the user. However, the social value of these exchanges of information is debatable, since they mostly give rise to competition and stigmatization. Additionally, the use of a MATE system is expected to increase an operator's confidence in the use of the machine or robot, thus making him or her less ashamed and more prone to interact with colleagues.

The roboethics ROBETH-R8 requirement demonstrates that, for professional users, difficulties with accessing robotic technologies may stem from a lack of confidence or knowledge of robotics \cite{Nevejans_2016}. The MATE approach aims at reducing these barriers by relying on an inclusive design. The inclusive design requirement is intrinsically satisfied with the implementation of a MATE system.

ROBETH-R9, the final roboethics requirement, has a marginal application to MATE systems. This requirement refers to the fact that, by using advanced robotic solutions, humans might lose the perception of their actual abilities and feel empowered by attributing a robot's skills to themselves \cite{Nevejans_2016}. The first step of a MATE system is the measurement of a human's skills, and, although stigmatization may be prevented, operators will unavoidably be faced with their own limitations measured by the system, as discussed in roboethics requirement ELSI-R6.

\begin{table*}
	\caption{\label{tab:users_needs}Users needs derived from the analysis of the limitations on the current human–machine systems.}
	\centering
	\renewcommand\arraystretch{1.05}
	\begin{tabular}{|l|}
		\hline
		\bf{1) Inclusion of all users}\\
		1.1) The system should be effectively usable by inexperienced operators\\
		1.2) The system should be effectively usable by operators with different age\\ 
		1.3) The system should be effectively usable by operators with different level of work experience or education\\
		1.4) The system should be effectively usable by operators with physical impairments\\
		\hline
		\bf{2) User-oriented organization of information}\\
		2.1) Procedures should adapt to the operator's skills\\
		2.2) The system should provide guided procedures for ordinary operations\\
		2.3) The system should guide the operator according to common practice solutions\\
		2.4) Specific prior training and studying the manual should not be necessary\\
		2.5) Operations should be performed in the correct sequence, according to the manual\\
		2.6) The system should suggest the operator what parameters need to be changed, based on the desired result\\
		\hline
		\bf{3) Prioritization of human factors}\\
		3.1) The system should be comfortable for all the users\\
		3.2) The stress level during the use of the system should be low\\
		3.3) The intervention of supervisors to assist the operators should be avoided\\
		3.4) Operators should feel confident when using the system\\
		\hline
		\bf{4) Enhancement of operator's performance}\\
		4.1) The number of errors should be reduced\\
		4.2) The execution time should be improved\\
		4.3) The correct operational mode and the correct value for critical parameters should be automatically selected\\
		4.4) The choice of wrong options should be prevented\\
		4.5) The HMI should depict the actual equipment and state of the machine\\
		\hline
		\bf{5) Advanced technological solutions}\\
		5.1) Hands-free interaction should be possible\\
		5.2) Portable interfaces should be available, to guide the operators in the working area\\
		\hline
	\end{tabular}
\end{table*}

\section{Impact of the MATE Approach in the Organization of a Company}\label{sec:questionnaires}
The practical implementation of the requirements in Table \ref{tab:design_recommendations} and the impact of a MATE system on the current organization of a company are examined in this section. Satisfying the general recommendations previously discussed requires that the design of new robotic or automated solutions follows the technical requirements and the principles of inclusive design; these may be applied differently depending on the specific characteristics of any given usage. Conversely, it is necessary for some concrete actions to be taken at a broader level that consider a company's organizational structure and goals. An infrastructure is necessary to support the introduction of a MATE system, as well as to preserve the operator's interests.

To address this issue, we explored how the ELSI analysis affects a company's organization. In particular, the intersection between the MEESTAR approach and the legal and social requirements leads to the following ELSI dimensions, which are discussed in detail in \cite{Sabattini_2017_CASE}:
\begin{itemize}
	\item occupational health
	\item occupational safety
	\item data protection
	\item ergonomic workplace design
	\item equal opportunities
	\item reintegration.
\end{itemize}
These factors represent the main organizational dimensions that are affected by the introduction of a MATE system in an industrial setting.

We first analyzed how companies are currently organized with respect to these factors, aiming to understand the corresponding successful measures previously taken and companies' predisposition to invest in these soft skills. We then estimated how a MATE system would affect these dimensions. The related potential for improvement and the associated risks following the introduction of a MATE interaction approach are discussed in \cite{Sabattini_2017_CASE}.

Information was gathered from a questionnaire that was designed to investigate the appropriateness of identified ELSI factors of MATE systems that were accessible to user groups with special needs and requirements. The questionnaire was organized in two parts: the first part investigated the status quo of the company with regard to the ELSI dimensions identified; the second part estimated the impact of a MATE system on these dimensions. The questionnaire was distributed to 13 company managers from the following diverse sectors: information technology and software development (two companies located in Germany), technology transfer (two located in Italy), industrial automation (four in Italy, one in Greece), white goods (one in Italy), packaging and bottling (two in Italy), and machine manufacturing (one in Germany).

\subsection{Status Quo of Companies with Regard to ELSI Dimensions}
\begin{figure}
	\centering
	\includegraphics[width=\columnwidth]{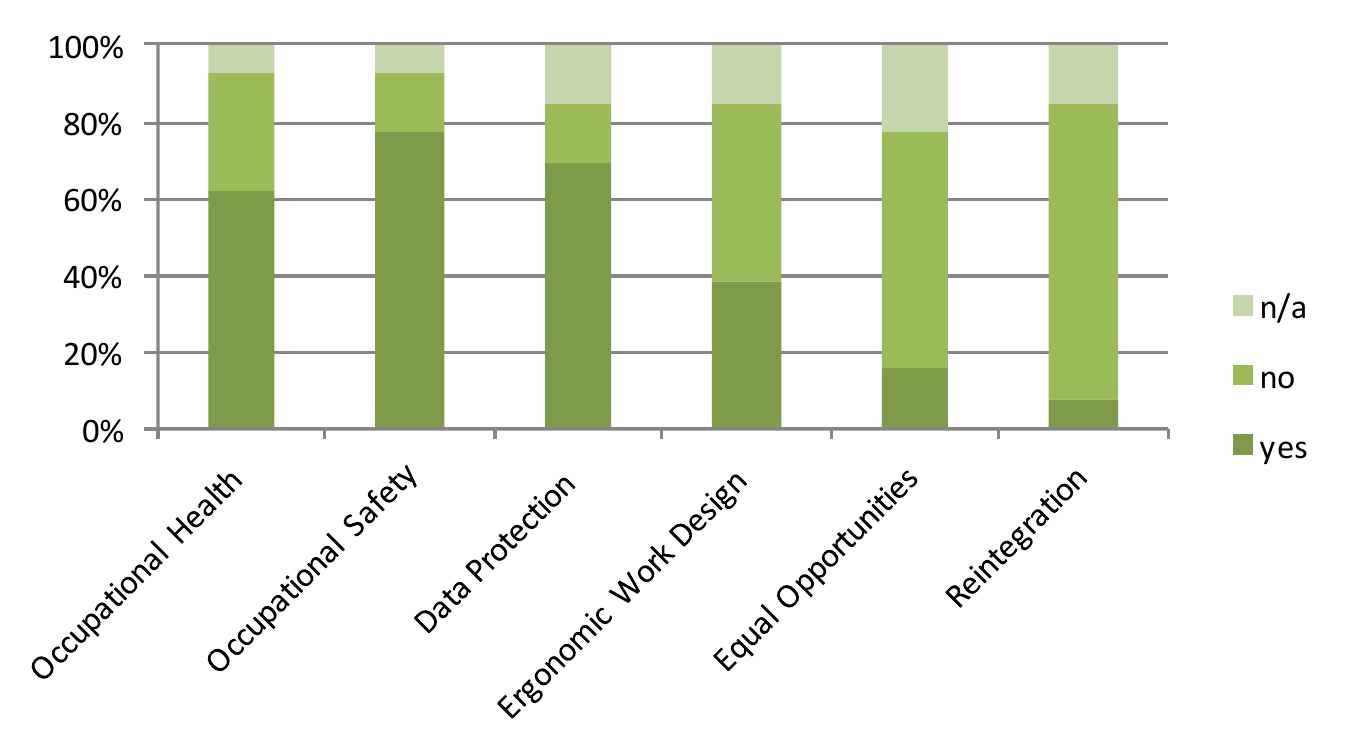}
	\caption{\label{fig:fig2_4_deliverable}The number of companies currently using a management system for each of the ELSI dimensions.}
\end{figure}

Of those that completed the questionnaire, Figure~\ref{fig:fig2_4_deliverable} illustrates an overview of the companies that currently include a management system for each of the ELSI dimensions. The data show that, among participants, the most commonly implemented management systems are occupational safety and health management systems and data protection divisions. It was demonstrated that, on average, companies used to employ more people in these divisions.

\begin{figure}
	\centering
	\includegraphics[width=1\columnwidth]{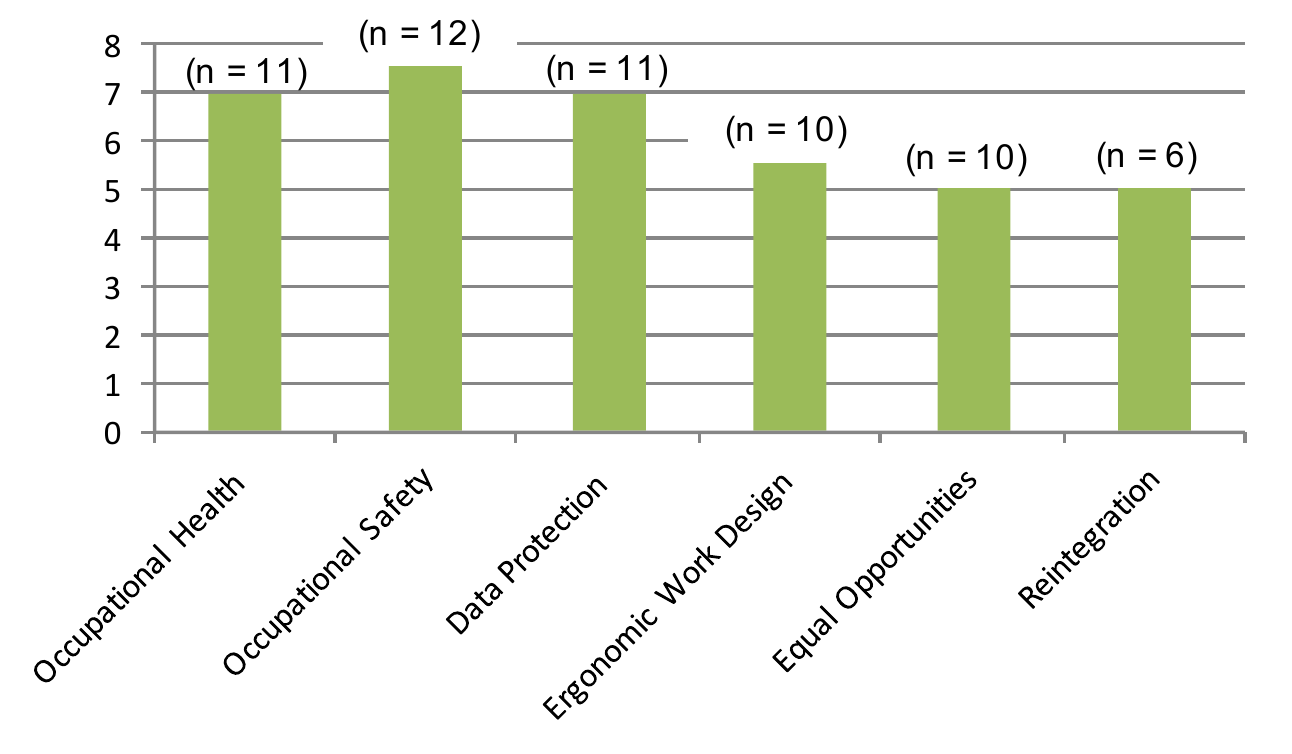}
	\caption{\label{fig:fig2_6_deliverable}The mean of the estimated overall relevance of ELSI dimensions by companies involved in the questionnaire. The number of averaged answers is reported above each bar.}
\end{figure}

Additionally, Figure~\ref{fig:fig2_6_deliverable} shows the mean value of the estimated overall relevance of each dimension for the companies involved in the questionnaire. Relevance was rated on a scale from 1 (very negative) to 10 (very positive). Occupational safety, with an average relevance of $7.6\pm2.7$, data protection at $7.0\pm2.5$, and occupational health at $6.9\pm2.9$, were found to be most relevant. The relevance of the other factors was rated at below mean: $5.6\pm3.3$ for ergonomic workplace design, $5.2\pm2.7$ for equal opportunities, and $5.2\pm3.0$ for reintegration. Note that, since some companies do not implement a management system for some of the ELSI dimensions examined, it was not possible to assess the corresponding relevance for all companies. Accordingly, the results in Figure~\ref{fig:fig2_6_deliverable} were averaged based on a different number of study participants. Furthermore, answers provided in the questionnaires indicated that successful measures previously implemented by the companies with respect to the ELSI factors included periodic health monitoring, psychological support, internal physiotherapists, fitness programs, reintegration of handicapped people, and reintegration after long-term illness.

\subsection{MATE Impact on ELSI Dimensions}
The second part of the questionnaire considered a working scenario where affective computing is applied to an industrial human–machine system, thereby measuring an operator's mental workload, stress, and induced anxiety by recording certain physiological signals. The questionnaire included specific questions regarding the following MATE scenario:
``\textit{The working machines are equipped with sensors that are able to track the strain of a working person by real-time measurement of his or her physiological parameters, e.g., heart rate and blood pressure. If the measured strain indicators are too high, the system adapts to the situation resulting in a lower stress level.}''

\begin{figure}
	\centering
	\includegraphics[width=1\columnwidth]{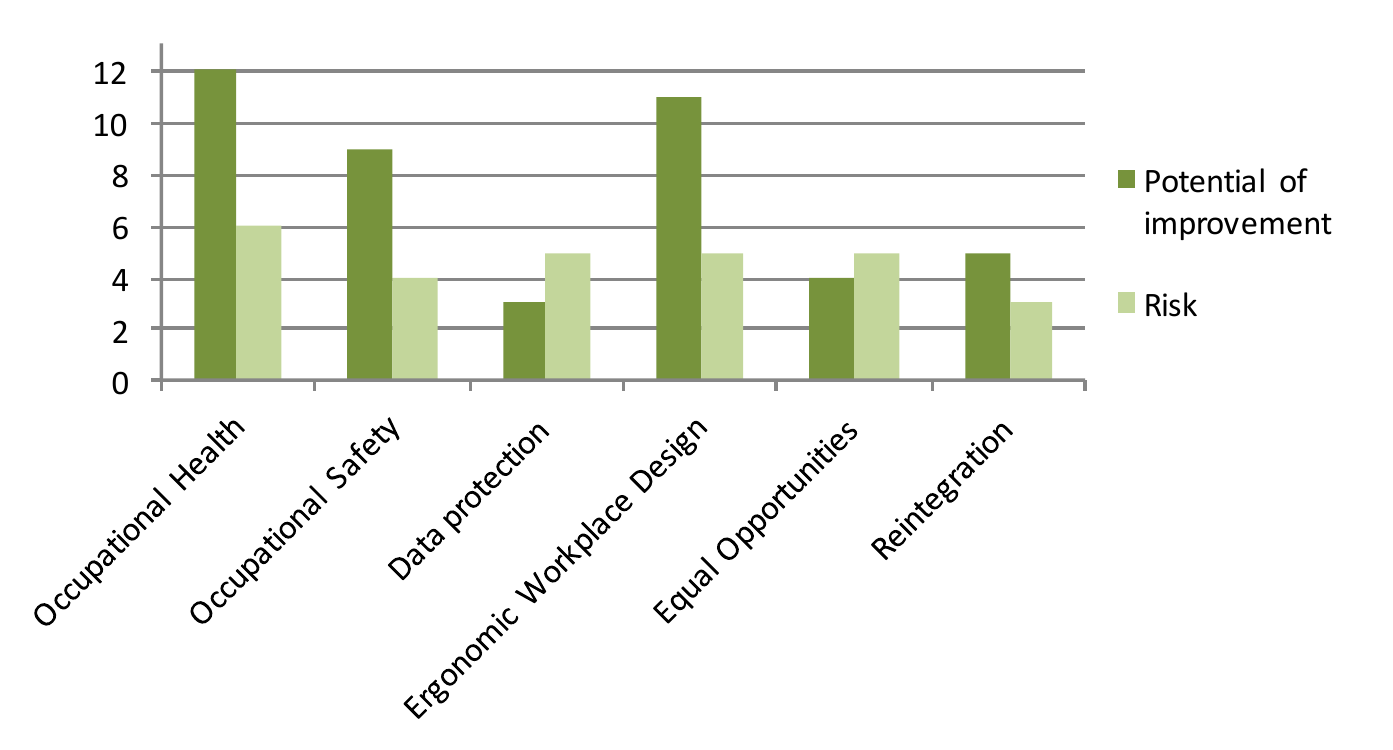}
	\caption{\label{fig:fig2_7_deliverable}The potential improvement and potential risks reported for each dimension of the ELSI concept.}
\end{figure}

Participants were first asked whether the potential for improvement or risks in measuring strain of a working person as described in the aforementioned scenario could be found (the results are summarized in Figure~\ref{fig:fig2_7_deliverable}). Most subjects reported the potential for improvement by means of the system in question: occupational health, 12 out of 13; ergonomic workplace design, 11 out of 13; and occupational safety, 9 out of 13. Occupational health, however, was also rated with the highest potential risks (six mentions), while five subjects out of 11 mentioned a potential risk with respect to data protection, ergonomic workplace design, and equal opportunities (two did not answer the corresponding questions). For data protection and equal opportunities, the estimated risk (five mentions each) overcomes the potential of improvement (three and four mentions, respectively); thus, these findings confirm that the potential discrimination of vulnerable users is a critical issue that needs to be addressed in the use of MATE systems, as described in the ELSI-R4 requirement. Data protection reported the smallest number of mentions for potential for improvement, which is clearly explained by the mission of a MATE system. Risks for occupational safety were mentioned by four users, and the smallest risks (three mentions) were reported for reintegration.

\begin{figure}
	\centering
	\includegraphics[width=1\columnwidth]{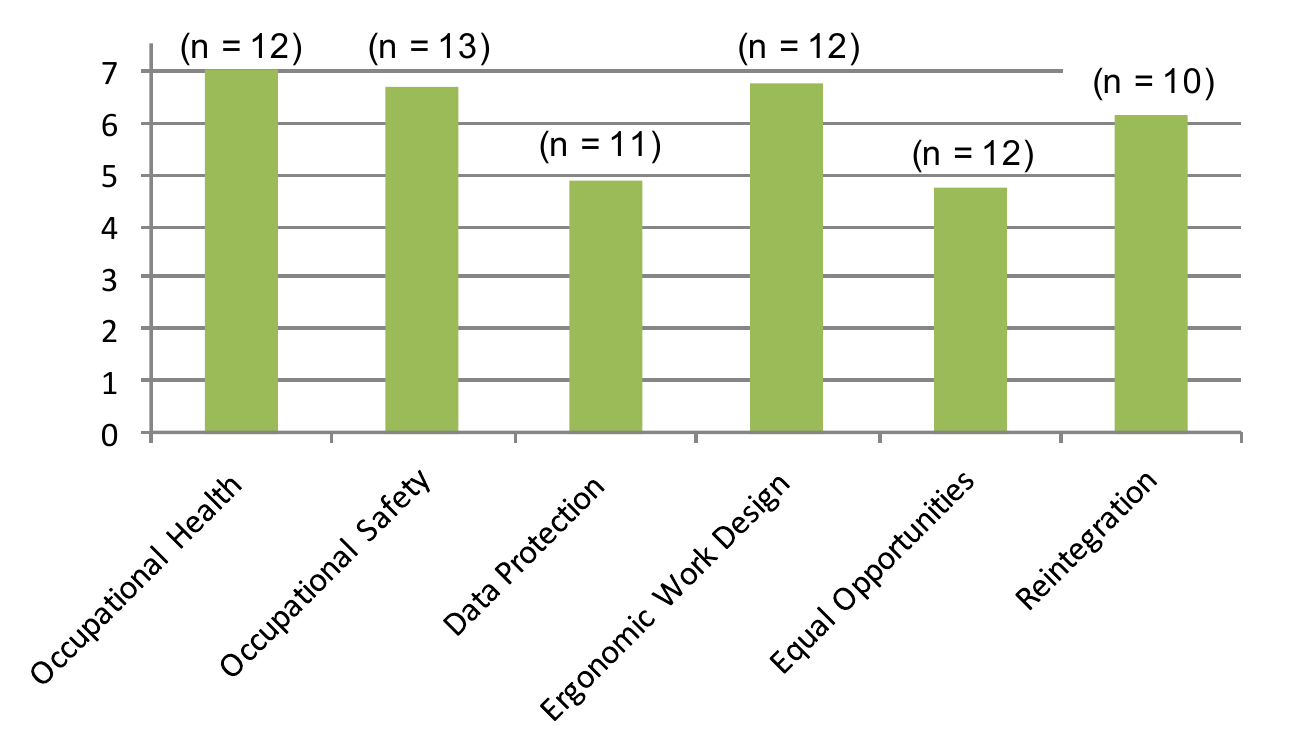}
	\caption{\label{fig:ELSI_positive_assessment}The mean impact of strain measurement per dimension of the ELSI concept. The number of averaged answers is reported above each bar.}
\end{figure}

Moreover, each participant was asked how positively the impact of strain measurement was considered for each dimension of the ELSI concept. Results, measured on a scale from one (very negative) to ten (very positive), are reported in Figure~\ref{fig:ELSI_positive_assessment}. The questionnaire responses indicated that the impact of strain measurements according to each dimension was mostly rated positively (above six points), with the exception of two dimensions: data protection, $4.9\pm1.9$; and equal opportunities, $4.8\pm2.0$. The influence of strain measurements on occupational health, $7.0\pm1.4$, was rated highest. The impact of stress measurements on occupational safety was rated $6.7\pm2.6$, and on ergonomic workplace design it was rated $6.8\pm2.2$; the impact on reintegration was rated $6.3\pm2.0$.

\section{Discussion and Conclusion}
In this article, we propose a new human-centered approach based on affective robotics and computing, with advanced automatic or robotic system designs. We call the resulting system \emph{MATE}. Measuring the operator's skill, performance, and stamina, it then adapts itself to facilitate a specific interaction. Teaching and support are provided to the operator to train him or her on the skills necessary to implement the system effectively. This type of system is intended to be usable for all users —particularly the most vulnerable ones— thus satisfying the requirements of inclusive design.

The social and ethical implications related to the use of MATE systems are reported in this article, and we describe the requirements that are essential to design a MATE system. Technical requirements are summarized and discussed in tandem with the ethical, legal, and social recommendations and principles inspired by roboethics.

Several organizational factors that are affected by the use of a MATE system are identified along with their impact on a company's structure. We then investigate how companies are currently organized in terms of such factors and what measures were previously taken to implement them.

Results indicated that more than 50\% of the participating companies implement occupational health, safety, and data protection management systems; however, there is still a lack of implementation in ergonomic workplace design, equal opportunities, or reintegration, though the estimated relevance for all ELSI dimensions was rated positively. Conversely, Figure~\ref{fig:fig2_6_deliverable} also demonstrates that relevance was rated highest for those ELSI dimensions that are currently in practice and well known. Since the MATE approach attempts to support the elderly, the impaired, and the low-skilled worker during machine and robotics operations, it thereby has the potential to create changes in the awareness of equal opportunities and reintegration for companies in the future. This assumption is consistent with regard to the estimated impact, which was rated high for health, safety, ergonomics, and reintegration.

The impact on data protection was rated surprisingly low, as shown in Figure~\ref{fig:ELSI_positive_assessment}, although potential risks were expected to be greater than the potential for improvement, as shown in Figure~\ref{fig:fig2_7_deliverable}. Equal opportunities play only a minor role compared to the other dimensions. The greatest potential for improvement is seen in health, safety, and ergonomics, which is consistent with the general objective of MATE systems. All of these dimensions were rated as having relatively low risks for the implementation of a MATE system.

With regard to the dimensions that are infrequently implemented, i.e., equal opportunities and reintegration, these were rated as having the lowest potential of improvement and were perceived to have the highest level of risk when compared to the potential for improvement. Therefore, a MATE system can not only demonstrate better awareness for ELSI factors, as mentioned previously, but it can also be mindful of specific dimensions that may have a greater level of risk and less potential for improvement when implemented in an industrial environment.

Lastly, the reliability of these results must be verified, and these issues should be examined further in subsequent investigations. Specifically, the proposed design recommendations and requirements will be reconsidered by assessing the real impact on concrete use cases brought by the MATE systems built in the framework of the INCLUSIVE project. The anticipated risks and improvements and possible side effects of strain measurement will be verified and updated, since possible other risks might arise, leading to further countermeasures to be considered in the design process.

Results show that more than 50\% of the participating companies implement occupational health, safety and data protection management systems, but that there is still a lack of implementation in ergonomic workplace design, equal opportunities or reintegration, though estimated relevance for all ELSI dimensions was rated positively. On the other hand, Fig.~\ref{fig:fig2_6_deliverable} also shows that relevance was rated highest for those ELSI dimensions that are already implemented in practice and well known. Since the here discussed \MATE{} approach aims at supporting elderly, impaired and unskilled workers during machine and robotics operations, it thereby has the potential to create changes in the awareness of equal opportunities and reintegration in the companies in the future. This assumption consolidates with regard to the estimated impact, which was rated high for health, safety, ergonomics and reintegration. 

The impact on data protection was rated surprisingly low, as shown in Fig.~\ref{fig:ELSI_positive_assessment}, although potential risks were expected to be grater than improvement potential, as shown in Fig.~\ref{fig:fig2_7_deliverable}. Also equal opportunities only plays a minor role compared to the other dimensions. 
Moreover, the highest potential of improvement is seen in health, safety and ergonomics, which is the general objective of \MATE{} systems. All these dimensions were rated with relatively low risks for the implementation of a \MATE{} system.

As regards the currently less implemented dimensions, namely equal opportunities and reintegration, they were rated with the lowest potential of improvement and with the highest risk if compared to potential improvement. Hence, a \MATE{} system cannot only retrieve better awareness for ELSI questions, as concluded earlier, but also should consider to pay special attention towards dimensions that might have higher risk than potential of improvement, when implemented in industrial environment.

Finally, the reliability of these results has to be verified and these issues have to be inspected more in detail in further investigations. Specifically, the proposed design recommendations and requirements will be reconsidered by assessing the real impact on concrete use cases brought by the \MATE{} systems built in the framework of the INCLUSIVE project. The foreseen risks and improvements and possible side effects of strain measurement will be verified and updated, since possible other risks might arise, thus leading to further counter measures to be considered in the design process.

%
%

\section*{Acknowledgement}
The research is carried out within the ``Smart and adaptive interfaces for INCLUSIVE work environment'' project, funded by the European Union's Horizon 2020 Research and Innovation Programme under grant agreement N. 723373. The authors would like to express their gratitude for the support given.

\bibliographystyle{IEEEtran}
\bibliography{biblio_SI_RAM}

\end{document}